\documentclass{llncs}
\usepackage{url}
\usepackage{graphicx}
\usepackage{color}
\newcommand{\comm}[1]{}
\begin{document}
\pagestyle{headings}  
\title{A Proof of Concept for Optimizing Task Parallelism by Locality Queues}
%
%
\author{Markus Wittmann\and Georg Hager}
\authorrunning{M. Wittmann et al.}   
%
%
\institute{Erlangen Regional Computing Center (RRZE),\\
University of Erlangen-Nuremberg,
Martensstra\ss e 1,  91058 Erlangen, Germany\\
\email{georg.hager@rrze.uni-erlangen.de}}

\maketitle              

\begin{abstract}
  Task parallelism as employed by the OpenMP \texttt{task} 
  construct, although ideal for tackling
  irregular problems or typical producer/consumer schemes, bears some
  potential for performance bottlenecks if locality of data access is
  important, which is typically the case for memory-bound code on
  ccNUMA systems.  We present a programming technique which
  ameliorates adverse effects of dynamic task distribution by sorting
  tasks into locality queues, each of which is preferably processed by
  threads that belong to the same locality domain. Dynamic scheduling
  is fully preserved inside each domain, and is preferred over
  possible load imbalance even if non-local access is required. The
  effectiveness of the approach is demonstrated using a blocked
  six-point stencil solver as a toy model.
\end{abstract}
\section{Dynamic Scheduling on ccNUMA Systems}
Dynamic scheduling is the preferred method for solving load imbalance
problems with shared memory parallelization. The OpenMP standard
provides the \verb.dynamic. and \verb.guided. schedulings for
worksharing loops, and the \verb.task. construct for task-based
parallelism.  
If the additional overhead for
dynamic scheduling is negligible for the application at hand, these
approaches are ideal on UMA (Uniform Memory Access) systems like the
now outdated single-core multi-socket SMP nodes, or multi-core chips
with ``isotropic'' caches, i.\,e.\ where each cache level is either
exclusive to one core or shared among all cores on a chip.

If, however, data access locality is important for performance and
scalability, dynamic scheduling of any kind is usually ruled out.
The most prominent example are memory-bound applications on ccNUMA-type
systems: Even if memory pages are carefully placed into the 
NUMA domains by parallel first-touch initialization, peak memory
bandwidth cannot be reached if cores access the NUMA domains
in a random manner, although this is still far better than 
serial initialization if there are no other choices. 

As a simple benchmark we choose a 3D six-point Jacobi solver 
with constant coefficients as recently studied extensively
by Datta et al.\ \cite{datta08}\@. 
The site update function,
\begin{eqnarray*}
F_{t+1}(i,j,k) & = & c_1F_{t}(i,j,k)\\
 &   &  {}+c_2\left[F_t(i-1,j,k)+F_t(i+1,j,k)+F_t(i,j-1,k)\right.\\
 &   &  {}+\left. F_t(i,j+1,k)+F_t(i,j,k-1)+F_t(i,j,k+1)
\right]~,
\end{eqnarray*}
is evaluated for each lattice site in a 3D loop nest,
and the memory layout is chosen so that $i$ is the fast
index.
Each site update (in the following called ``LUP'') incurs 
seven loads and one store, of which, at large problem sizes,
one load and one store 
cause main memory traffic if suitable spatial blocking is applied.
This leads to a code balance of 3 bytes per flop (assuming
that non-temporal stores are not used so that a store miss
causes a cache line read for ownership), so the code is clearly
memory-bound on all current cache-based architectures.
In what follows we use a problem size of $600^3$ sites 
($\approx$\,3.25\,GiB of memory for double precision variables) 
and a block size of
600$\times$10$\times$10 ($d_i${}$\times${}$d_j${}$\times${}$d_k$) 
sites, unless otherwise noted. The update loop nest iterates
over all blocks in turn, and 
standard worksharing loop parallelization is done over the outer ($k$-blocking) loop (initialization is performed via the identical scheme):
\begin{verbatim}
#pragma omp parallel for schedule(runtime)
  for(int kb=0; kb<number_of_k_blocks; ++kb) {
    for(int jb=0; jb<number_of_j_blocks; ++jb) {
      for(int ib=0; ib<number_of_i_blocks; ++ib) {
        jacobi_sweep_block(ib,jb,kb);
  } } }
\end{verbatim}
Note that with the standard \verb.i. block size equal to
the extent of the lattice in that direction (which is required
to make best use of the hardware prefetching capabilities
on the processors used), \verb.number_of_i_blocks.
is equal to one.
The \verb.jacobi_sweep_block(). function performs one
Jacobi sweep, i.e.\ one update per lattice site, over
all sites in the block determined by its parameters.
In case of dynamic loop scheduling,
parallel first-touch initialization is done via a 
\verb.static,1. (round robin) loop schedule, whereas
plain \verb.static. scheduling is used otherwise.

\begin{figure}[tbp]
\centering
\includegraphics*[width=0.75\textwidth]{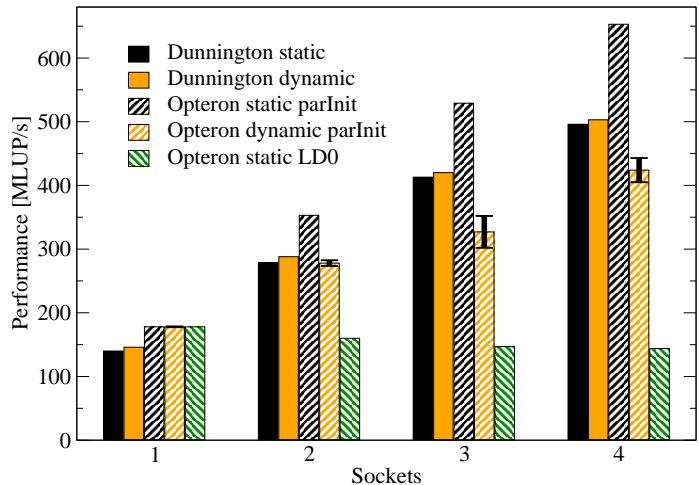}
\caption{\label{fig:umavsnuma}Peformance in million lattice site
    updates per second (MLUP/s) versus
    number of sockets for an OpenMP parallel 6-point stencil solver
    on a UMA (solid bars) and a ccNUMA (hatched bars) system,
    using standard worksharing loop parallelization(see text for details).
    ``parInit'' denotes parallel first-touch data initialization.
    The ``LD0'' data set was obtained by forcing all memory pages
    to reside in locality domain 0\@.}
\end{figure}

Fig.~\ref{fig:umavsnuma} illustrates the impact of dynamic
scheduling on the solver's scalability for two benchmark systems:
\begin{itemize}
\item ``Dunnington'' is an EA Intel UMA server system (``Caneland'' chipset)
    with four six-core Intel Xeon 7460  processor chips at
    2.66\,GHz. Data for this system is included for illustrative
    purposes.
\item ``Opteron'' is an HP DL585 G5 ccNUMA server with four locality
    domains (LDs), one per socket, and four dual-core AMD Opteron 
    8220 SE processor chips at 2.8\,GHz. The processors are
    connected via HyperTransport 1.0\,GHz links (4\,GB/s per
    direction)\@.
\end{itemize}
Both systems ran current Linux kernels, and the Intel C++ compiler,
version 11.0.074 was used for all benchmarks.
As we are mostly interested in scalability data, detailed
performance characteristics for the platforms under consideration
are omitted.
One should note, however, that there is significant optimization
potential in stencil codes like the one we use here. The block size 
we have chosen is close to optimal from a data transfer 
perspective \cite{datta08},
and the performance data obtained is in line with STREAM COPY
scalability on the same systems.\comm{mit eintragen?} 
In all cases, the number of threads per socket was chosen so that
the local memory bus could be saturated, which happens to be the case for
two threads on both platforms. Core-thread affinity was enforced
by overloading the \verb.pthread_create(). call and using
\verb.sched_setaffinity(). in turn for each newly created thread,
skipping the OpenMP shepherd thread(s)\@.\comm{tbb?}

\begin{figure}[tbp]
\centering
\includegraphics*[width=0.75\textwidth]{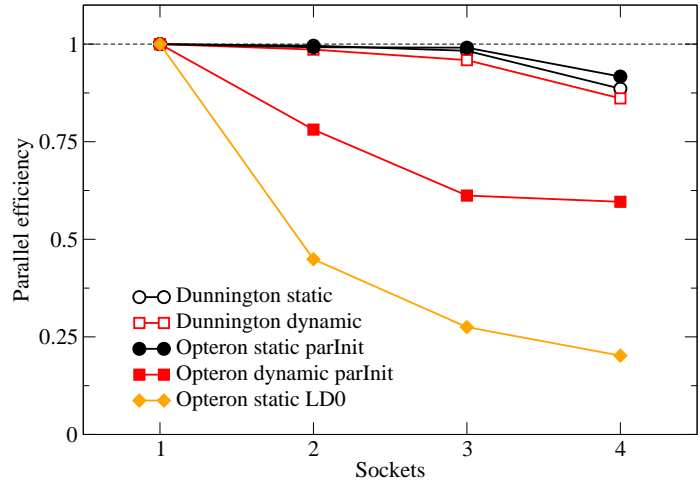}
\caption{\label{fig:umavsnuma-e}Parallel efficiency $\varepsilon$ versus
    number of sockets for the same data sets as in 
    Fig.~\ref{fig:umavsnuma}\@.} 
\end{figure}
The performance results and parallel efficiency numbers in 
Figs.~\ref{fig:umavsnuma} and \ref{fig:umavsnuma-e} show that dynamic scheduling
has negligible impact on the UMA system for the chosen problem and
block sizes, although one may of course expect a noticeable performance
hit if OpenMP startup and scheduling overhead become dominant with
small data sets and block sizes. If static scheduling and proper
parallel initialization are employed, the ccNUMA system shows similar
characteristics as the UMA node (bandwidth scalability with four sockets is 
not ideal for current Opteron-based systems because of protocol
overhead)\@. Dynamic scheduling, however, has a catastrophic effect
on parallel efficiency as remote accesses and contention on the HyperTransport
network dominate performance. Moreover, there is a noticeable
statistical performance variation because access patterns vary
from sweep to sweep. Nevertheless,
due to the round-robin page placement as described above, there
is still some parallelism available. If we force all memory pages
to be mapped in the first locality domain (LD0), all parallelism
is lost and performance is limited by the single-domain memory
bandwidth, which is already saturated with the two local threads.
Execution is hence serialized.


\section{Tasking and Locality Queues on ccNUMA}

In contrast to standard worksharing loop parallelization, tasking
in OpenMP requires to split the problem into a number of work ``packages'',
called \emph{tasks}, each of which must be submitted to an internal
pool by the \verb.omp task. directive. For the Jacobi solver we define
one task to be a single block of the size specified above. This is
in contrast to loop worksharing where one parallelized outer 
loop iteration consisted of all blocks with the same \verb.kb. 
coordinate. Using the \verb.collapse. clause on the parallel loop
nest would correct this discrepancy, but there is no further insight
gained.

\subsection{Standard Tasking}\label{sec:tasking}

The tasks are produced (submitted) by a single thread and consumed 
by all threads and in a 3D loop nest:
\begin{verbatim}
#pragma omp parallel
{
  #pragma omp single
  {
    for(int kb=0; kb<number_of_k_blocks; ++kb) {
      for(int jb=0; jb<number_of_j_blocks; ++jb) { 
        for(int ib=0; ib<number_of_i_blocks; ++ib)  {
          #pragma omp task
            jacobi_sweep_block(ib,jb,kb);
    } } }
  }
}
\end{verbatim}
Submitting the tasks in parallel is possible but did not make
any difference in the parameter ranges considered here (but see
below for the impact of submission order)\@. There is still a choice
as to how first-touch initialization should be performed, so we compare
\verb.static. and \verb.static,1. scheduling for loop initialization.
\begin{table}[tbp]
\renewcommand{\arraystretch}{1.3}\setlength{\tabcolsep}{6pt}
\caption{\label{tab:tvsq}Performance of the Jacobi solver in MLUP/s
    on 8 threads of
    the Opteron (ccNUMA) platform with two different schedulings for
    the block initialization loop (rows), comparing standard tasking 
    and tasking with locality queues, and the two possible choices
    for the submit loop nest order.}
\begin{center}
\begin{tabular}{l||c|c|c|c}\hline
     & \multicolumn{2}{c|}{\bfseries tasking} & \multicolumn{2}{c}{\bfseries tasking + queues} \\\hline\hline
\bfseries submit order &  \ttfamily kji   &   \ttfamily jki   &   \ttfamily kji   &  \ttfamily jki \\\hline
\bfseries \texttt{static} init   & $149.8\pm0.2$  &   $247.9\pm 0.6$ &   $180.8\pm 0.4$  & $598.2\pm 2.9$ \\
\bfseries \texttt{static,1} init & $205.9\pm 0.4$ &   $412.7\pm2.8$  &            $588.4\pm 3.1$  & $594.6\pm 4.2$ \\\hline
\end{tabular}
\end{center}
\end{table}
Table \ref{tab:tvsq} shows performance results (columns labeled ``tasking'')
on the ccNUMA platform, using eight threads and two different loop 
orderings for the submission loop. With static initialization and
the standard \verb.kji. loop
order as shown above, performance is roughly equal to the results
obtained in the previous section with LD0 enforcement (``Opteron
static LD0'' data
in Figs.~\ref{fig:umavsnuma} and \ref{fig:umavsnuma-e}), i.e.\ 
execution is serialized. Performance is slightly improved to
roughly the 4-thread dynamic scheduling level (``Opteron
dynamic parInit'' data for two sockets in Fig.~\ref{fig:umavsnuma}) 
by choosing the \verb.jki. loop order for submission. Going to
\verb.static,1. initialization, the 8-thread dynamic scheduling
performance can be matched.

The large impact of submit and initialization orders can be explained
by assuming that there is only a limited number of ``queued'', i.e.\
unprocessed tasks allowed at any time. In the course of executing the 
submission loop, this limit is reached very quickly and the
submitting thread is used for processing tasks for some time.
From our measurements, the limit is equal to 257 tasks with
the compiler used. One \verb.ib.-\verb.jb. layer of our grid comprises
60 tasks (with the chosen problem and block sizes), and 
60 layers are available, which amounts to 3600 tasks in total.
With static scheduling, one block of 257 consecutive tasks is
usually associated with a single locality domain (rarely two), hence
the serialization of memory access. Choosing \verb.static,1. 
scheduling for initialization, each consecutive layer is placed 
into a different locality domain, but 257 tasks comprise only
slightly more than four layers. Assuming that the order of execution
for tasks resembles \verb.static,1. loop workshare scheduling 
because each thread is served a task in turn, the number of LDs
to be accessed in parallel is limited (although it is hard to predict
the actual level of parallelism)\@. Finally, by choosing the
\verb.jki. submission loop order, consecutive tasks cycle through
locality domains, and parallelism is as expected from dynamic
loop scheduling. The statistical variation is surprisingly small,
however.

These observations document that it is nontrivial to employ 
tasking on ccNUMA systems and reach at least the performance level
of standard dynamic loop scheduling. In the next section we
will demonstrate how task scheduling under 
locality constraints can be optimized by substituting
part of the OpenMP scheduler by user program logic. 

\subsection{Tasking with locality queues}\label{sec:queues}

Each task,
which equals one lattice block (or tile) in our case, is associated
with a C++ object (of type \verb.BlockObject.) and equipped
with an integer locality variable which denotes the locality 
domain it was placed in upon initialization. The submission loop
now takes the following form:
\begin{verbatim}
#pragma omp parallel
{
  #pragma omp single
  {
    for(int kb=0; kb<number_of_k_blocks; ++kb) {
      for(int jb=0; jb<number_of_j_blocks; ++jb) { 
        for(int ib=0; ib<number_of_i_blocks; ++ib)  {
          BlockObject p(kb,jb,ib);
          queues[p.locality()].Enqueue(p);
          #pragma omp task
            process_block_from_queue(queues);
    } } }
  }
}
\end{verbatim}
The \verb.queues. object is basically a \verb.std::vector<>. of \verb.std::queue<>.
objects, each associated with one locality domain, and each
protected from concurrent access with an OpenMP lock. Calling the
\verb.Enqueue(). method of a queue appends a block object to it. As shown above,
blocks are sorted into those \emph{locality queues} according to their
respective locality variables. One OpenMP task, executed by
the \verb.process_block_from_queue(). function, now consists of two parts:
\begin{enumerate}
\item Figuring out which LD the executing thread belongs to
\item Dequeuing the oldest waiting block in the locality queue belonging 
    to this domain and calling \verb.jacobi_sweep_block(). for it
\end{enumerate}
If the local queue of a thread is empty, other queues are tried
in a spin loop until a block is found:
\begin{verbatim}
void process_block_from_queue(LocalityQueues &queues) {
  // ...
  bool found=false;
  BlockObject p;
  int ld = ld_ID[omp_get_thread_num()];
  while (!found) {
    found = queues[LD].Dequeue(p);
    if (!found) {
      ld = (ld + 1) % num_of_lds;
    }
  }
  jacobi_sweep_block(p.ib,p.jb,p.kb);
}
\end{verbatim}
The global \verb.ld_ID. vector must be preset with a correct 
thread-to-LD mapping. Enqueuing and dequeuing blocks
from a queue is made thread-safe by protecting the queue with an OpenMP
lock. 

Note that scanning other queues if a thread's local queue is empty
gives load balancing priority over strict access locality, which may or
may not be desirable depending on the application. The team of threads
in one locality domain shares one queue, so scheduling is purely
dynamic in that case. 

There is actually a ``race condition'' with the described scheme because 
it is possible that some task executes a block just queued 
before the corresponding task is actually submitted. This is 
however not a problem because the number of submitted tasks is
always equal to the number of queued blocks, and no task will ever be left
waiting for new blocks forever.

Table \ref{tab:tvsq} shows performance results for eight threads
with four locality queues under the columns labeled ``tasking + queues''\@. 
For \verb.static. initialization and the \verb.kji. submission order, 
the limited overall number of waiting tasks has the same consequences as 
with plain tasking (see Sect.~\ref{sec:tasking})\@. In this case,
although the queuing mechanism is in effect, a single queue holds most
of the tasks at any point in time. All threads are served from this queue
and thus execute in a single LD\@. However, using
the alternate \verb.jki. submission order or \verb.static,1. 
initialization, all queues are fed in parallel and threads can 
be served tasks from their local queue. Performance then comes
close to static scheduling within a 10\,\% margin (see 
Fig.~\ref{fig:umavsnuma})\@.

One should note that a similar effect could have been achieved with
nested parallelism, using one thread per LD in the outer parallel
region and several threads (one per core) in the nested region.
However, we believe our approach to be more powerful and easier 
to apply if properly wrapped into C++ logic that takes care of 
affinity and work distribution. Moreover, the thread pooling strategies
employed by many current compilers inhibit sensible
affinity mechanisms when using nested OpenMP constructs.

\section{Summary and outlook}

We have demonstrated how locality queues can be employed to optimize 
parallel memory access on ccNUMA systems when OpenMP tasking is used.
Locality queues substitute the uncontrolled, dynamic task scheduling
by a static and a dynamic part. The latter is mostly restricted to
the cores in one NUMA domain, providing full dynamic load balancing
on the LD level. Scheduling between domains is static, but load balancing
can be given priority over strictly local access. The larger the number of 
threads per LD, the more dynamic the task distribution, so our scheme
will get more interesting in view of future many-core processors.

Using locality queues to optimize real applications on ccNUMA systems
is under investigation, as well as a study of possible additional 
overhead of the method for ``small'' tasks. The same methodology
may also be applied for optimizing \verb.parallel_while. constructs
in Intel Threading Building Blocks (TBB, \cite{tbb})\@.
Further potentials, not restricted to ccNUMA architectures, may be found in 
the possibility to implement temporal blocking (doing more than
one time step on a block to reduce pressure on the memory subsystem
\cite{nitsure}) by associating one locality queue to a number of cores
that share a cache level. As an advantage over static temporal blocking,
no frequent global barriers would be required.\vspace*{1cm}

\noindent{\bfseries\large Acknowledgments}\vspace*{0.3cm}

\noindent Fruitful discussions with Gerhard Wellein and Thomas
Zeiser are gratefully acknowledged.

%
%

\end{document}